\documentclass[twocolumn]{openjournal}

\usepackage{graphicx} 
\usepackage{multirow}
\usepackage{amsmath}
\usepackage{booktabs}
\usepackage[colorlinks,linkcolor=blue,citecolor=blue,urlcolor=blue ]{hyperref}
\usepackage[utf8]{inputenc}
\usepackage{float}
\usepackage{xcolor}
\usepackage{ulem}
\usepackage[T1]{fontenc}
\newcommand{\RNum}[1]{\uppercase\expandafter{\romannumeral #1\relax}}
\usepackage{caption}
\usepackage{orcidlink}
\newcommand{\kms}{\,\ensuremath{\rm{km\,s}^{-1}}}
\newcommand{\msun}{\mbox{${\rm M}_\odot$}}
\newcommand{\mstar}{\mbox{${M}_{\rm star}$}}
\newcommand{\ha}{\mbox{${\rm H}\alpha$}}
\newcommand{\ewr}{\mbox{$W_{\rm r}(2796)$}}

\newcommand{\RN}[1]{%
  \textup{\uppercase\expandafter{\romannumeral#1}}%
}
\newcommand{\abs}[1]{\left\lvert#1\right\rvert}

\begin{document}
\title{On the connection between galaxy orientation and halo absorption properties}

\author{Rohan Venkat$^{1}$}
\thanks{Corresponding authors: Rohan Venkat and Hsiao-Wen Chen} \email{rvenkat@uchicago.edu, hwchen@uchicago.edu}
\author{Soo May Wee$^{1}$}
\author{Hsiao-Wen Chen$^{2,3}$\orcidlink{0000-0001-8813-4182}}
\affiliation{$^{1}$Department of Physics, The University of Chicago, Chicago, IL 60637, USA}
\affiliation{$^{2}$Department of Astronomy and Astrophysics, The University of Chicago, Chicago, IL 60637, USA}
\affiliation{$^{3}$Kavli Institute for Cosmological Physics, The University of Chicago, Chicago, IL 60637, USA}


\begin{abstract}
We present a systematic investigation of the azimuthal dependence of metal-line absorption in the circumgalactic medium (CGM) using a uniformly selected sample of 87 isolated galaxies at $z < 0.4$ from the Magellan MagE \ion{Mg}{2} (M3) halo survey. High-quality archival imaging enables quantitative morphological measurements---including disk inclination and position angle---for every galaxy, providing a robust framework for assessing how absorber strength depends on the geometric alignment between galaxies and the QSO sightlines. All galaxies have associated constraints on \ion{Mg}{2}$\lambda\,2796$ absorption, and a subset of 56 galaxies also have measurements of \ion{Ca}{2}$\lambda\,3934$. We compare rest-frame \ion{Mg}{2} and \ion{Ca}{2} equivalent widths with both projected distance and deprojected galactocentric distance. Across the full sample, we find no statistically significant correlation between absorption strength and azimuthal angle. Restricting to the 71 galaxies with well-determined disk orientations reveals an apparent excess of strong \ion{Mg}{2} absorbers near the projected major axis, but a Kendall's $\tau$ test confirms that this trend is not statistically significant. \ion{Ca}{2} absorption, which exhibits a low covering fraction of $\kappa_{\rm CaII} = 0.18^{+0.06}_{-0.04}$ within 50 kpc for $W_r(3934) > 0.1$ \AA, shows no measurable azimuthal dependence.  To assess potential biases, we quantify the effects of projection, disk inclination, and variations in imaging quality. After accounting for these systematics, the spatial distribution of low-redshift \ion{Mg}{2} and \ion{Ca}{2} absorbers is consistent with arising from a randomly distributed population, with no compelling evidence for azimuthal anisotropy at $d \lesssim 50$ kpc. A larger sample with robust constraints on the disk orientation will be required to uncover or rule out subtle anisotropic trends.
\end{abstract}

\maketitle
\section{Introduction}
\label{sec:intro}

Understanding the distribution and motions of gas within galaxies and their surrounding environments has remained a primary interest in astrophysics research. Star formation is fueled by gas; therefore, constraining how galaxies acquire gas and subsequently drive outflows is essential to understanding their growth and evolution history. The circumgalactic medium (CGM) resides in the interface between star-forming regions and the low-density intergalactic medium \citep[e.g.,][]{Chen_Zahedy_2025} and contains a large fraction of the gas content in a galaxy \citep[e.g.,][]{Li_2018}, as well as heavy elements synthesized in stars \citep[e.g.,][]{Peeples_2014}. The CGM serves as a primary source of gas to sustain star formation in galaxies. 

Examination of circumgalactic gas in simulated galaxies at low redshift suggests a complex circulation of outflowing and inflowing gas characterized by distinctly non-spherical geometry \citep[e.g.,][]{Shen2013, Schneider2020}.  Starburst-driven winds from nearby galaxies erupt perpendicular to the galactic disk in conical lobes with opening angles of $\theta\approx 10-45\deg$ \citep[e.g.,][]{Veilleux2005} because this direction marks the largest gradient in the gas density \citep[e.g.,][]{deYoung1994}.  In contrast, accretion of intergalactic gas is expected to proceed along the disk plane with $\lesssim 10$\% covering fraction on the sky \citep[e.g.,][]{CAFG2011, Fumagalli2011}.  Such an expected azimuthal dependence of the spatial distribution of infalling and outflowing gas has spurred a strong community interest in exploring the presence or absence of correlation between metal-line absorbers observed along the transverse direction from a foreground galaxy and its disk orientation \citep[e.g.,][]{Bordoloi_2011, Bouche2012, Kacprzak2015, Peroux2016, Martin_2019}.

Specifically, an initial study by \citet{Bordoloi_2011} showed that at projected distances of $d < 50$ physical kpc (pkpc) from $z\approx 0.7$ star-forming galaxies, the mean \ion{Mg}{2}\,$\lambda\lambda\,2796, 2803$ absorption is roughly twice as strong along the minor axis as within 45$^\circ$ of the major axis \citep[see also][]{Lan2018}. Subsequently, \citet{Bouche2012} reported a bimodal distribution of disk–quasar alignments out to $d\approx 90$ physical kpc, although the sample was small, containing 10 \ion{Mg}{2}-absorbing galaxies.  \citet{Kacprzak2015} extended these analyses to highly-ionized \ion{O}{6}-bearing gas and found a similar bimodality out to $d\approx 200$ pkpc.  Such bimodality is qualitatively consistent with a picture in which metal-bearing gas observed along the minor axis traces biconical outflows, whereas absorption along the major axis arises from inflows in or near the disk plane. 

\begin{figure*}
    \centering
    \includegraphics[width=0.95\textwidth]{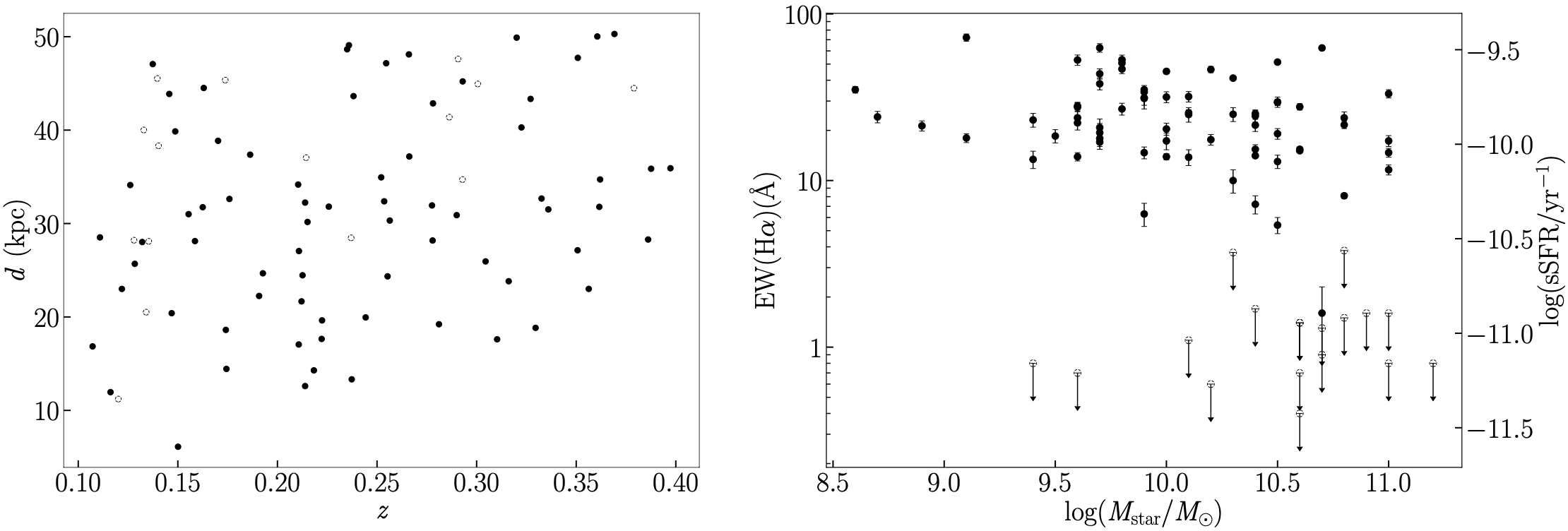}
    \caption{Summary of galaxy properties for the adopted sample. Panel (\textit{a}) displays the projected distance ($d$) versus redshift $z$, while panel (\textit{b}) displays the distribution of H$\alpha$ equivalent width (EW) relative to stellar mass, \mstar.  Open symbols in panel (\textit{a}) are galaxies with no \ion{Mg}{2} absorption detected to a sensitive upper limit (see \S\,\ref{sec:sample}), while galaxies with no detected H$\alpha$ emission are shown as downward arrows in panel (\textit{b}) to mark 2-$\sigma$ upper limits on the underlying H$\alpha$ emission flux.  As a proxy for characterizing the star formation history, the observed EW(H$\alpha$) is converted to specific star formation rate (sSFR) following \cite{Huang2021} and displayed on the right of panel (\textit{b}).}
    \label{fig:galaxy_summary}
\end{figure*}

However, the situation is more complicated. Comparing CGM and ISM metallicities, \citet{Peroux2016} found that minor-axis sightlines are significantly more metal-deficient than the host ISM, seemingly at odds with the expectation of chemically enriched outflows \citep[cf.,][]{Peroux2020}.  In addition, using a sample of 50 star-forming galaxies at $z\approx 0.2$, \citet{Martin_2019} reported enhanced \ion{Mg}{2} absorption along the major axis out to $d\approx 40$ pkpc (see their Figure 11), in stark contrast to earlier findings.  

Such discrepancies may be attributed to intrinsic differences between different galaxy samples from different cosmic epochs.  Alternatively, uncertainties in measuring galaxy orientation, as well as ambiguities in distinguishing between co-planar and minor-axis gas around moderately inclined galaxies due to projection effects, may also lead to discrepant azimuthal dependence.  

To evaluate the impact of sample selection, uncertainties in galaxy orientations, and ambiguities from moderate disk inclinations, we performed a quantitative morphological study of galaxies with available constraints on the CGM \ion{Mg}{2} absorption strength. 
Specifically, we utilize the galaxy and absorber sample from the Magellan MagE \ion{Mg}{2} (M3) halo project \citep[][]{Chen2010a, Huang2021}.  The M3 halo project is designed to establish a statistically representative sample of $z\lesssim 0.4$ galaxies with known \ion{Mg}{2} absorption constraints, to determine the incidence, spatial extent, and covering fraction of \ion{Mg}{2}-bearing gas surrounding galaxies of different properties.  All galaxy--QSO pairs from the M3 Halo project were chosen without prior knowledge of the presence or absence of absorbing gas to ensure an unbiased characterization of the metal-enriched CGM.  The focus on low-redshift systems enables robust morphological analysis using archival ground-based imaging data and a detailed investigation of the correlation — or lack thereof — between galaxy orientation and halo absorption.

The paper is organized as follows.  In Section \ref{sec:sample}, we describe the galaxy sample and available imaging and spectroscopic data.  In Section \ref{sec:method}, we describe the morphological analysis and resulting empirical quantities and associated uncertainties. In Section \ref{sec:discussion}, we discuss the implications of our findings.  Throughout the paper, all distance measurements are in physical units under a standard $\Lambda$ cosmology, $\Omega_{\rm M} = 0.3$, $\Omega_\Lambda=0.7$, and a Hubble constant $H_{\rm 0} = 70\rm \,km\,s^{-1}\,Mpc^{-1}$.

\section{The Galaxy Sample and data}
\label{sec:sample}

From the M3 halo project, we identify a sample of 97 galaxies within $d\le 50$ pkpc.  Existing \ion{Mg}{2} absorption constraints from \cite{Chen2010a} and \cite{Huang2021} enable us to test the galaxy--\ion{Mg}{2} connection in the inner CGM, motivated by the discrepant reports of azimuthal bimodality at $d\lesssim 45$ kpc between the findings of \citet{Bordoloi_2011} and \citet{Martin_2019}.  Of the 97 galaxies, 10 are found to have close neighbors (within a projected distance less than the size of the gaseous halo $R_{\rm gas}$ and a line-of-sight velocity separation $|\Delta v| < 300$ \kms; see \citealt{Chen2010a, Huang2021}) based on existing spectroscopic survey data.  Because such multiplicity obscures a clear azimuthal assignment for the absorber, we exclude these galaxies, leaving a total of 87 isolated galaxies for the morphological studies.  A summary of the sample characteristics is presented in Figure \ref{fig:galaxy_summary} with the projected distance $d$ versus redshift distribution displayed in the left panel and \ha\ equivalent width, ${\rm EW}(\ha)$, versus galaxy stellar mass, \mstar, displayed in the right panel. We place a 2-$\sigma$ upper limit on EW(H$\alpha$) for galaxies with no detectable \ha\ emission \citep[see][for details]{Huang2021}.

As described in \citet{Huang2021}, ${\rm EW}(\ha)$ serves as a proxy of the specific star formation rate (sSFR) given that it is the ratio of total \ha\ line flux, tracing SFR over the last 10 Myr of the star formation history \citep[e.g.,][]{Kennicutt2012}\footnote{The timescale probed by H$\alpha$ emission is substantially shorter than the timescale required to transport chemically enriched material to galactocentric distances of tens of kpc \citep[e.g.,][]{Suchkov:1994}. Consequently, the presence or absence of a correlation between CGM metal absorption and H$\alpha$-inferred star formation activity should not be interpreted as direct evidence for or against a causal connection. Instead, any such correlation is likely influenced by the cumulative star-formation and feedback history of a galaxy over much longer timescales.}, and continuum flux, tracing total stellar mass.  We adopt the correlation between ${\rm EW}(\ha)$ and sSFR from \cite{Belfiore2018}. The corresponding sSFR is indicated in the right axis of Figure \ref{fig:galaxy_summary}.  In Table \ref{tab:galaxies}, we provide for each galaxy its ID, redshift, $z_{\rm gal}$, angular separation from the background QSO, $\Delta\theta$, the corresponding projected distance, $d$, stellar mass, \mstar, the observed H$\alpha$ equivalent width, EW(H$\alpha$), and inferred halo radius, $R_h$, in columns (1) through (7). 

For the isolated sample of 87 galaxies, we adopt the rest-frame \ion{Mg}{2} equivalent widths and associated uncertainties reported by \citet{Huang2021}. These measurements are listed in column (12) of Table 1, with non-detections assigned 2-$\sigma$ upper limits derived from the noise properties of the QSO spectra. The same QSO spectra also provide coverage of the \ion{Ca}{2},$\lambda\lambda$3934,3969 doublet for 56 galaxies in the sample. Calcium is known to be heavily depleted onto dust grains---more than 99\% of Ca resides in the solid phase \citep[e.g.,][]{Savage:1996, Wild:2006}---and its low ionization potential, combined with this high depletion fraction, implies an underlying large neutral hydrogen column density \citep[e.g.,][]{Wakker:2000}, as commonly found in star-forming disks and in conical outflows. We therefore include the \ion{Ca}{2} absorption strength in our assessment of possible azimuthal dependence. The rest-frame \ion{Ca}{2} equivalent widths and uncertainties for these 56 galaxies are provided in column (13) of Table 1, and---as with \ion{Mg}{2}---2-$\sigma$ upper limits are assigned to non-detections based on the noise characteristics of the QSO spectra.

\begin{figure*}
    \centering
    \includegraphics[width=\linewidth]{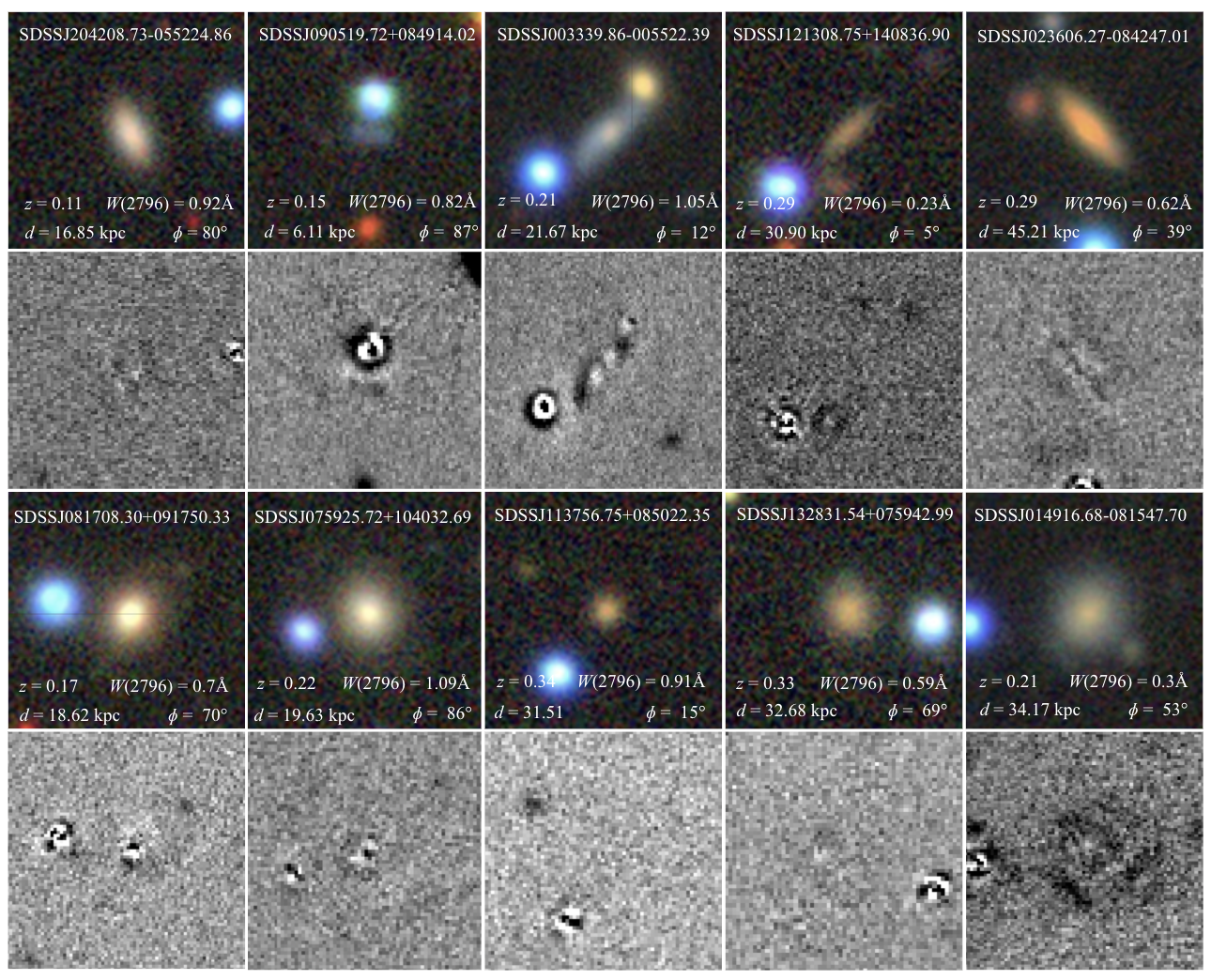}
    \caption{Examples of galaxies in the dataset taken from DECaLs Sky Browser feature ordered by impact parameter from $d = 14.4$ kpc to $d = 47.7$ kpc from left to right and top to bottom. The images are 30 arcsecs on a side and are centered on the galaxies with the associated quasar being the adjacent bright blue object. Included in the images are the redshift, \ion{Mg}{2} absorption strength, impact parameter, and the best-fit orientation angle $\phi$ of the galaxy's major axis relative to the QSO sightline. Note that the residuals are constructed by removing nearby bright galaxies from the input image in addition to the quasar light.}
    \label{fig:gal_grid}
\end{figure*}

Figure \ref{fig:galaxy_summary} shows that galaxies selected in our study are uniformly distributed in projected distance, from $d=6$ to $d=50$ kpc with a median of $\langle\,d\,\rangle=31.5$ kpc, and in redshift, from $z=0.11$ to $z=0.40$ with a median of $\langle\,z\,\rangle=0.23$.  Additionally, the sample encompasses a range of star formation histories typical of the field population \citep[e.g.,][]{Leja:2022}, including 17 galaxies with minimal ongoing star formation.  All galaxies have existing multi-bandpass $g$-, $r$-, and $z$-band images from the Dark Energy Camera Legacy Survey with 5-$\sigma$ imaging depths of $g=24$, $r=23.4$, and $z=22.5$ mag for an exponential disk of half-light radius $r_e=0.45\arcsec$ \citep[DECaLS;][]{Dey2019}.  Although the $g$-band imaging is typically $\approx 0.6$ mag deeper, it is more sensitive to young stellar populations.  In contrast, the $r$-band better traces the more evolved stellar populations and therefore provides a more representative view of the underlying galaxy structure.  To obtain robust morphological constraints for our intermediate-redshift galaxies, we therefore base our analysis on the $r$-band images retrieved from the DECaLS archive. The mean point spread function (PSF) full-width-at-half-maximum (FWHM) ranges from ${\rm FWHM}=0.9\arcsec$ to $1.4\arcsec$, corresponding to a spatial scale of $\approx 3.3$--5 kpc at the median redshift, $z=0.23$, of the galaxy sample.  As discussed in \S\,\ref{sec:seeing} below, at the relatively low redshifts of our galaxies, the image quality is sufficient for placing robust constraints on key morphological parameters, including the sizes, inclinations (axis ratio), and position angles of the stellar components. 


\section{Morphological Analysis}
\label{sec:method}

We employ \texttt{GALFIT} \citep[][]{GALFIT} to constrain the morphologies of all 87 galaxies in our sample.  As described in \S,\ref{sec:sample}, the imaging data are retrieved from the DECaLS archive together with the associated inverse-variance maps, which are subsequently converted to the corresponding noise image. The morphological analysis is performed using the $r$-band images, which offer an optimal balance between depth and image quality \citep[see Table~4 in][]{Dey2019}.

\texttt{GALFIT} requires as inputs an image of the galaxy–quasar pair, a point-spread function (PSF) image, and a corresponding sigma (noise) image. It returns the set of best-fit morphological parameters that reproduce the observed two-dimensional surface-brightness distribution for each object, along with a residual image obtained by subtracting the best-fit model from the data. Following standard practice, each galaxy is modeled with a PSF-convolved \citet{Sersic:1968} profile to account for the atmospheric seeing smoothing.  The expected intrinsic flux at pixel ($x$, $y$) for a S\'ersic profile is given by
\begin{equation}
\mu(x,y)=\mu_e\exp\left\{-b_n\left[r/r_e\right ]^{1/n}-1\right\},
\label{eq:sersic}
\end{equation}
with $b_n$ coupled to the S\'ersic index $n$ such that the flux enclosed within the effective radius $r_e$ equals half of the total galaxy flux.  The radial coordinate is defined as $r=\sqrt{(x-x_0)^2+(y-y_0)^2}$, where ($x_0$, $y_0$) marks the galaxy center.  To account for the apparent ellipticity of the two-dimensional isophotal structure, the model includes an axis ratio $b/a$, corresponding to an inclination angle, $i=\arccos(b/a)$, for disky galaxies, and a position angle PA describing the orientation of the major axis relative to the cardinal direction.  The morphological parameters provided by \texttt{GALFIT} for extended galaxy profiles are, therefore, $x_0$, $y_0$, $r_e$, $b_n$, $n$, $b/a$, and PA, in addition to a sky background.  This intrinsic S\'ersic model is then convolved with a Gaussian PSF with the FWHM determined empirically using stars in each field before being compared to the data.  As evident in Figure \ref{fig:gal_grid}, an accurate determination of the galaxy morphology requires simultaneous modeling of the QSO light for close galaxy-QSO pairs (e.g., SDSSJ090519.72$+$084914.02).  A point source model is, therefore, added for modeling the QSO light.

We experiment with two-component S\'ersic fits for all galaxies but find no significant improvement in the residuals, indicating that the current imaging data do not warrant additional model complexity. We therefore adopt single-component S\'ersic models for all galaxies. To illustrate the quality of the fits, Figure~\ref{fig:gal_grid} presents representative examples spanning a range of morphologies. Across the full sample of 87 galaxies, the S\'ersic indices have a median value of 0.86, with 16th and 84th percentiles of 0.5 and 1.8, respectively, and a tail extending to 6.6. 

Although \texttt{GALFIT} incorporates the noise map when estimating uncertainties for all morphological parameters, we find that the reported errors are often underestimated and that the best-fit parameters can be sensitive to the initial guesses provided by the user. Because accurate uncertainty estimates — particularly in the ellipticity and orientation of the major axis — are essential to our analysis, we developed a custom Monte Carlo fitting routine to obtain robust and statistically meaningful errors. Our custom code takes the input image and its corresponding noise map, then generates a series of synthetic realizations by perturbing each pixel value according to its uncertainty, assuming a Gaussian distribution. For each realization, we run \texttt{GALFIT} with randomly perturbed initial parameter guesses to derive a new set of best-fit morphological parameters. This process is repeated 1000 times, and the median and dispersion of each parameter distribution are adopted as the best-fit values and their associated uncertainties.


A key quantity in evaluating the presence or absence of an azimuthal dependence in the observed spatial distribution of metal-line absorbers around galaxies is the orientation of the galaxy's major axis relative to the line-of-sight of the background QSO.  We adopt the best-fit PA of the galaxy and the position of the background QSO relative to the galaxy, ($\delta_{qx}$, $\delta_{qy}$).  The azimuthal angle, $\phi$, can then be computed following
\begin{equation}
    \phi = \tan^{-1}\abs{\frac{1 + \frac{\delta_{qy}}{\delta_{qx}}\,{\rm tan}^{-1}({\rm PA})}{\frac{\delta_{qy}}{\delta_{qx}}+{\rm tan}^{-1}({\rm PA})}}
    \label{eq:phi}
\end{equation}
Due to the angular symmetry of a spiral galaxy, calculating the uncertainty in $\phi$ is a nontrivial task. We determine the range of $\phi$ that is consistent with our morphological estimates as follows. The estimated uncertainty in PA, $\Delta\text{PA}$, marks a 1-$\sigma$ lower and upper bounds in PA, denoted $\text{PA}_{\rm min}$ and $\text{PA}_{\rm max}$, respectively. The corresponding minimum (maximum) value of $\phi$, denoted $\phi_{\rm min}$ ($\phi_{\rm max}$), can be found by considering all PA's in the range [$\text{PA}_{\rm min}$,$\text{PA}_{\rm max}$] and taking the minimum (maximum) $\phi$ in this range. This procedure then gives for each galaxy a range [$\phi_{\rm min}$,$\phi_{\rm max}$] and consequently the corresponding uncertainties in $\phi$ based on the uncertainties in PA. 

Adopting the best-fit morphological parameters also allows us to compute a deprojected galactocentric distance $r_d$ corresponding to the observed projected distance, $d$, between the galaxy and the QSO on the sky, following
\begin{equation}
r_d = d \left[ 1 + \sin^2 (\phi) \tan^2(i) \right]^{1/2}.
    \label{eq:rd}
\end{equation}
This quantity is particularly relevant when considering a scenario in which the observed \ion{Mg}{2} absorption arises in an extended gaseous disk associated with the stellar component of the galaxy \citep[see e.g.,][]{Chen_1998}.  The best-fit morphological parameters and their associated uncertainties, including the effective radius, $r_e$, axis-ratio inferred inclination angle, $i$, and the azimuthal angle of the major axis, $\phi$, are reported in columns (8) through (10).  The deprojected galactocentric distance, $r_d$, is presented in column (11).


\begin{figure*}
    \centering
    \includegraphics[width=0.925\textwidth]{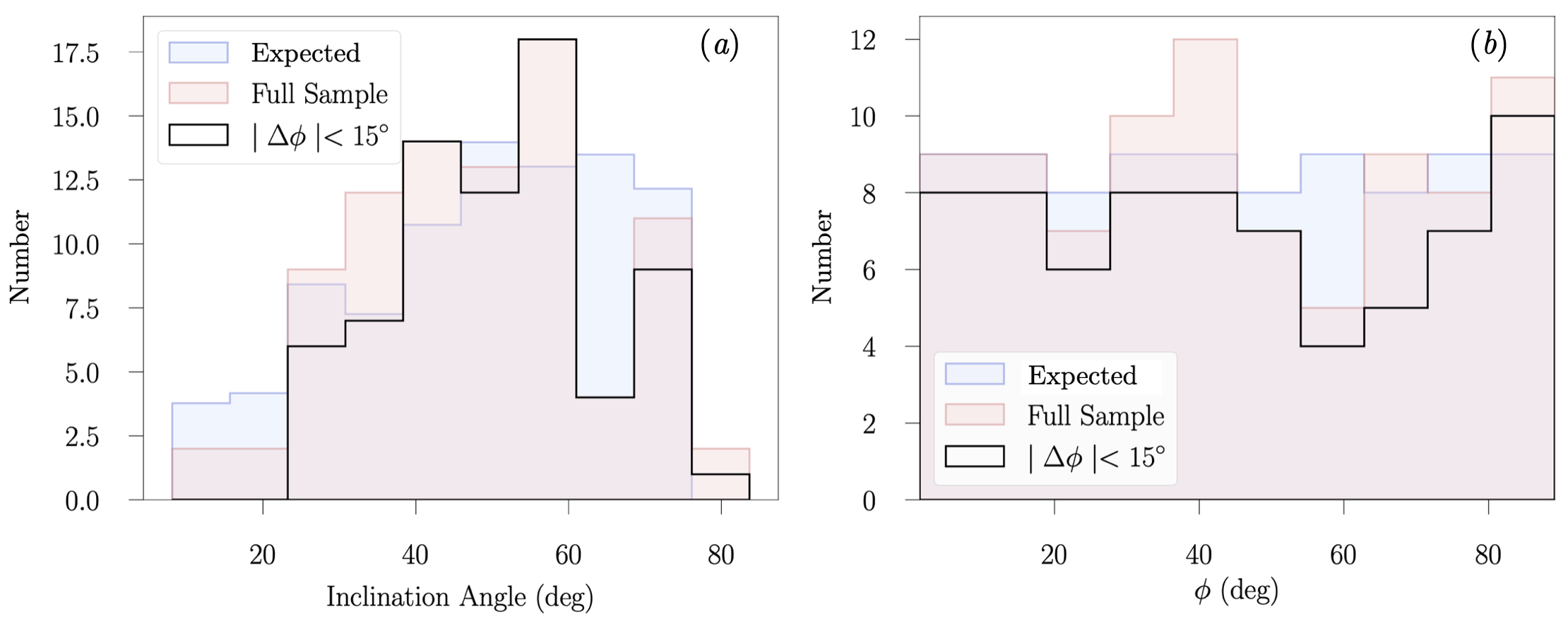}
    \caption{Summary of galaxy alignments relative to the background QSO sightlines. Panel (\textit{a}) shows the inclination-angle distribution inferred from galaxy axis ratios for the full sample (red) and for the subsample with well-constrained morphological parameters (black; defined as having orientation uncertainties $\Delta\,\phi<\!15^\circ$). Panel (b) shows the corresponding distribution of azimuthal angles, $\phi$. In both panels, the expected distributions for a randomly oriented disk population are shown in blue. The observed inclination and azimuthal-angle distributions are broadly consistent with expectations for a randomly oriented galaxy population. See the main text for details regarding the construction of the theoretical distributions.}
    \label{fig:phidistrth}
\end{figure*}


\section{Discussion and conclusions}
\label{sec:discussion}

We have conducted a detailed morphological analysis of a large sample of galaxies selected based on their proximity to background QSO sightlines. The galaxies are drawn from the M3 Halo Project \citep[e.g.,][]{Chen2010a, Huang2021} and comprise 97 systems at projected distances $d \leq 50$ kpc and redshifts $z < 0.4$. Of these, 10 galaxies have close companions and are excluded from the analysis, leaving 87 isolated galaxies with quantitative morphological measurements. This represents the largest galaxy sample with morphological constraints available in the literature. In the following sections, we examine the sample characteristics to assess potential selection biases and explore the relationship between the observed CGM absorption properties and galaxy morphology.
 
\subsection{Representativeness of the galaxy sample}
\label{sec:seeing}

Evaluating possible azimuthal dependencies in CGM absorption first requires determining whether our galaxy sample is representative of the broader disk-galaxy population. For an isotropic population, both inclination angles and azimuthal orientations should be randomly distributed with respect to any given line of sight. We therefore compare the observed distributions of (1) inclination angles inferred from galaxy axis ratios and (2) orientation angles relative to the QSO sightlines, against the expectations for a randomly oriented disk population. The comparison is performed for the full galaxy sample, irrespective of morphological-parameter uncertainties, and separately for a subsample of 71 galaxies with well-determined orientations, defined by uncertainties in the position-angle measurement of $\Delta \phi < 15^\circ$.

To generate the theoretical expectation for the inclination-angle distribution, $i$, we account for both the intrinsic probability of observing a galaxy at a given inclination and the effects of seeing, which impose a minimum spatial scale for resolving galaxy shapes. For an isotropically oriented disk population, we adopt a probability distribution function of $p(i) \propto \cos(i)$ for the intrinsic inclination distribution. The seeing constraint arises from the fact that smoothing suppresses reliable identification of small, highly inclined disks. To quantify this effect, we carry out Monte Carlo simulations in which we (i) generate synthetic galaxies with sizes of the major axis spanning the observed range, (ii) assign each galaxy an inclination drawn from the expected random distribution, and (iii) convolve the model images with the typical seeing in our data. The resulting inclination distribution is shown as the blue-shaded histogram in Figure \ref{fig:phidistrth}\textit{a}. The decline toward high inclinations reflects the intrinsic $\cos(i)$ dependence, while the suppression at low inclinations is driven by seeing. The model reproduces the observed distributions well for both the full sample (red histogram) and the subsample with well-constrained orientation angles (open black histogram).

Similarly, the observed and expected azimuthal-angle distributions are presented in Figure \ref{fig:phidistrth}\textit{b}. Although the full sample shows an apparent deficit near $\phi \approx 60^\circ$, this feature is diminished when the analysis is restricted to the 71 galaxies with well-constrained orientation measurements. Taken together, these comparisons indicate that the galaxy sample is statistically consistent with a randomly oriented disk population relative to the QSO sightlines. While we find no compelling evidence for preferential alignment, a larger sample with uniformly robust morphological constraints will be required to draw more definitive conclusions.

\begin{figure*}
    \centering
    \includegraphics[width=0.925\linewidth]{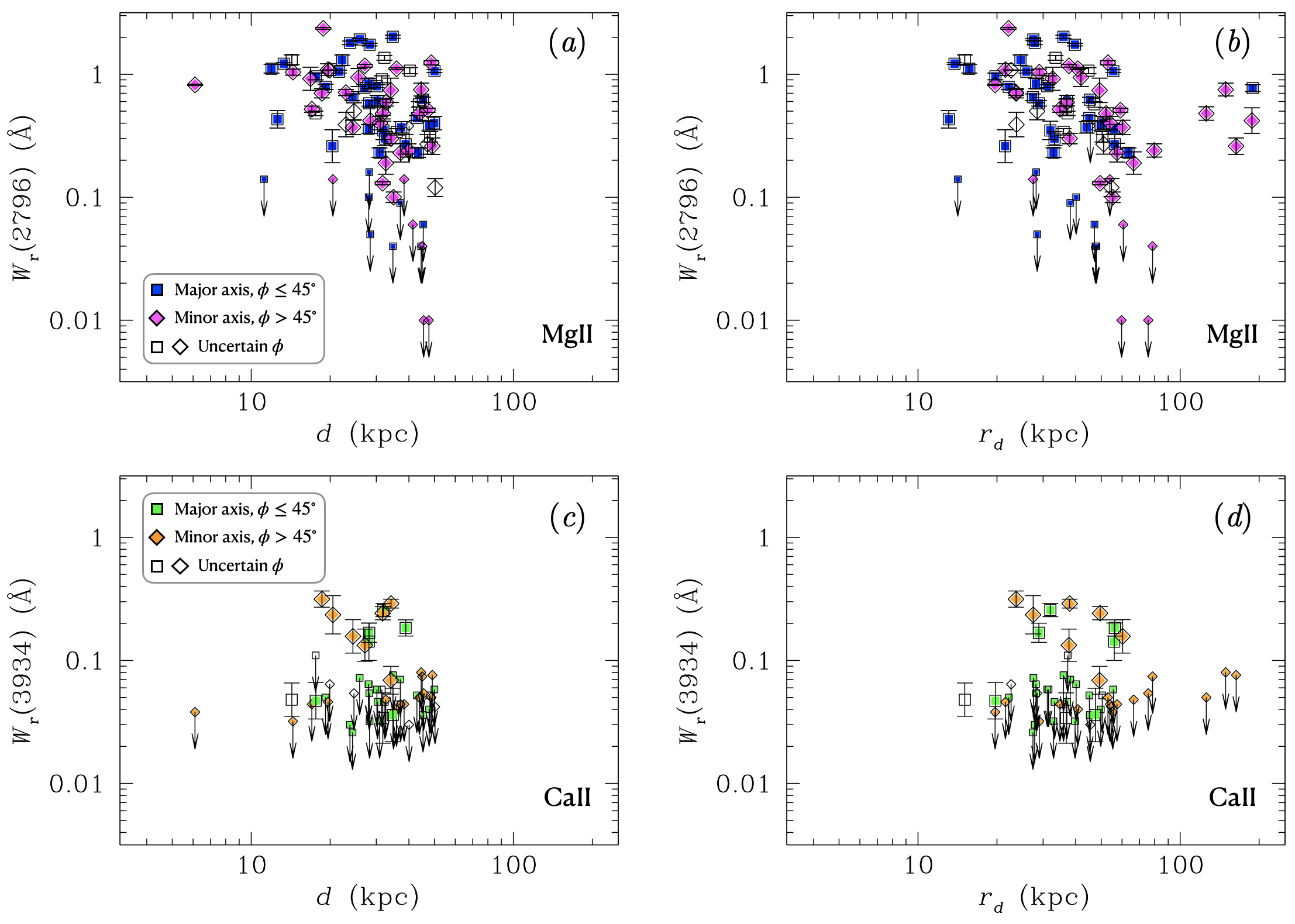}
    \caption{Observed \ion{Mg}{2} (\textit{top} panels) and \ion{Ca}{2} (\textit{bottom} panels) absorption strengths versus distance for galaxies in our sample.  The sample is divided into three subsamples based on their azimuthal constraints.  Galaxies whose long axis are reliably aligned closer to the QSO sightline ($\phi\leq 45^\circ$ and $|\Delta\phi|<15^\circ$) are shown as squares; those with reliably determined minor-axis orientation ($\phi>45^\circ$ and $|\Delta\phi|<15^\circ$) are shown as diamonds; systems with poorly constrained azimuthal angles ($|\Delta\phi|>15^\circ$) are shown as open symbols. The \textit{left} panels show the rest-frame absorption equivalent widths as a function of projected distance, $d$. The \textit{right} panels present the corresponding equivalent widths relative to the deprojected galactocentric distance, $r_d$, computed along the extended disk plane based on the best-fit inclination and orientation angles following Equation \ref{eq:rd}.}
    \label{fig:distr}
\end{figure*}

\subsection{Radial distribution of metal-line absorbers}
\label{sec:radial}

Next, we examine the radial distribution of metal-line absorption strengths to enable direct comparisons with previous CGM studies. A common observable in absorption-line analyses is the rest-frame equivalent width as a function of projected distance from the host galaxy. With constraints on disk inclination and orientation, we can further evaluate whether the absorbers trace extended gaseous disks associated with star-forming regions by adopting the deprojected galactocentric distance from Equation \ref{eq:rd} \citep[e.g.,][]{Chen_1998}.

Because our galaxy sample is drawn from the M3 survey---designed to characterize the \ion{Mg}{2}-bearing CGM---the radial behavior of \ion{Mg}{2} absorption and its dependence on galaxy mass and star formation history have already been thoroughly quantified in \citet{Chen2010a} and \citet{Huang2021}. In Figure \ref{fig:distr}\textit{a}, we reproduce the relation between the rest-frame \ion{Mg}{2} equivalent width, $W_r(2796)$, and projected distance $d$ for all galaxies at $d \lesssim 50$ kpc included in this study. For galaxies in which \ion{Mg}{2} absorption is not detected, we display 2$\sigma$ upper limits derived from the noise properties of the QSO spectra (shown as downward arrows). Figure \ref{fig:distr}\textit{b} presents the same measurements as a function of the de-projected galactocentric distance, $r_d$. Panels (\textit{c}) and (\textit{d}) show analogous relations for the \ion{Ca}{2} 3934 transition, plotting $W_r(3934)$ versus $d$ and $r_d$, respectively, for the 56 galaxies in our sample with available constraints on \ion{Ca}{2}. In all four panels, the sample is subdivided into three categories according to the quality of their azimuthal-angle constraints. Galaxies with reliably determined major-axis alignment toward the QSO sightline ($\phi \leq 45^\circ$ and $|\Delta\phi| < 15^\circ$) are shown as squares; those with reliably determined minor-axis alignment ($\phi > 45^\circ$ and $|\Delta\phi| < 15^\circ$) are shown as diamonds; and systems with poorly constrained azimuthal angles ($|\Delta\phi| > 15^\circ$) are shown as open symbols.

Three notable features are clear in Figure \ref{fig:distr}.  First, adopting the deprojected galactocentric distance yields a tighter inverse correlation between decreasing $W_r(2796)$ and increasing $r_d$, but with five distinct outliers at $r_d>100$ pkpc.  Four of the outliers are galaxies with QSO sightlines passing through regions closer to the minor axis, indicating that an extended disk model does not apply.  These five galaxies exhibit a range in ${\rm EW}(\ha)$, from ${\rm EW}(\ha)<0.6$ \AA\ to ${\rm EW}(\ha)\approx 28$ \AA.  The broad range of star-formation histories suggested by the observed ${\rm EW}(\ha)$ highlights the complexity of linking star-formation history to chemical enrichment in the CGM and the likelihood that multiple physical processes contribute to metal transport in galaxy halos. Secondly, such inverse-correlation between absorber strength and distance is not seen in \ion{Ca}{2}, either with $d$ or $r_d$ \citep[cf.,][]{Zhu:2013, Rubin:2022, Ng:2025}.  Finally, no significant difference has been found in the observed radial profile of either \ion{Mg}{2} or \ion{Ca}{2} between major- and minor-axis regions.  

However, the covering fraction of \ion{Ca}{2}-absorbing gas is substantially lower than that of \ion{Mg}{2}. Among 56 galaxies at $d<50$ kpc, only 10 show associated \ion{Ca}{2} absorption stronger than $W_r(3934)=0.1$ \AA, while the remaining systems exhibit no absorption with a 2-$\sigma$ upper limit that is either comparable to or below 0.1 \AA. These measurements yield a covering fraction of $\kappa_{\rm CaII}=0.18_{-0.04}^{+0.06}$ for \ion{Ca}{2} absorbers with $W_r(3934)> 0.1$ \AA\ within 50 kpc in halos around galaxies.  In contrast, previous studies report a much higher covering fraction, $\kappa_{\rm MgII} > 0.7$, for \ion{Mg}{2} absorbers with $W_r(2796)>0.1$ \AA\ out to 100 kpc \citep[e.g.,][]{Huang2021}. The significantly lower covering fraction of \ion{Ca}{2} results in a correspondingly smaller sample of \ion{Ca}{2}–galaxy pairs, highlighting the need for a larger dataset to strengthen the statistical constraints.

\begin{figure*}
    \centering
    \includegraphics[width=\linewidth]{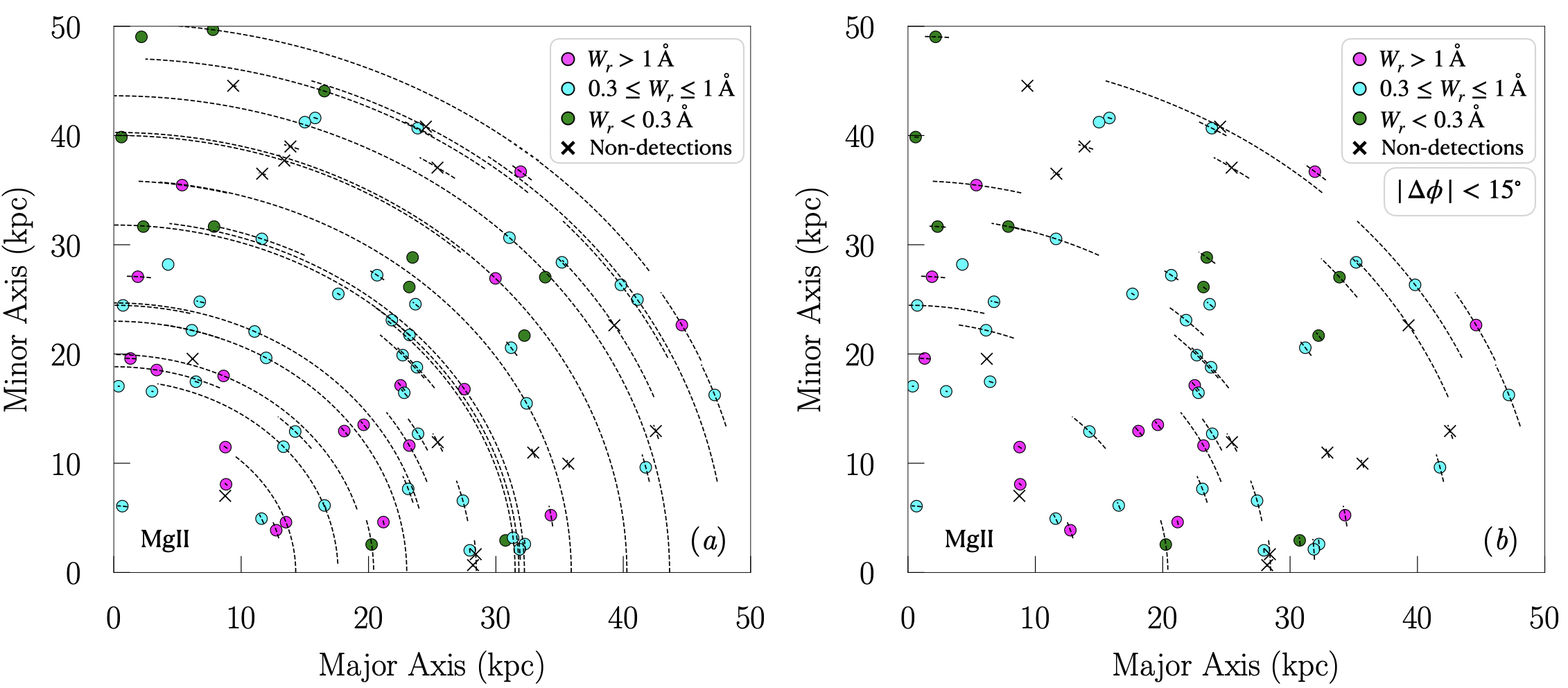}
    \includegraphics[width=\linewidth]{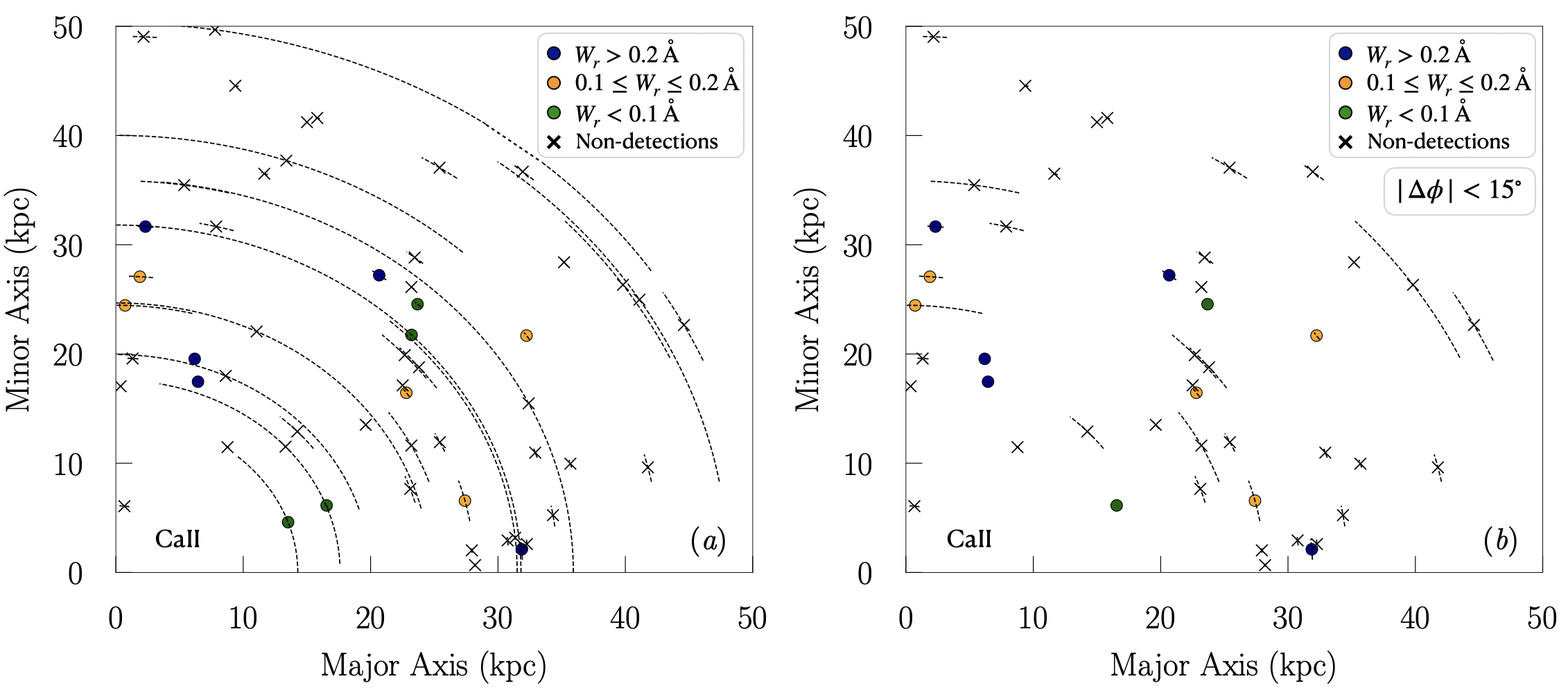}
    \caption{Observed variations in absorber strength as a function of galaxy geometric alignment relative to the QSO sightlines. Each panel shows the projected position of the QSO sightline relative to the semi-major and semi-minor axes in the rest-frame disk plane of the absorbing galaxy placed at (0,0) of the panel. Galaxies are grouped into four subsamples based on their rest-frame equivalent width. For \ion{Mg}{2} (\textit{top} row), systems are classified as strong ($W_r > 1$ \AA; magenta), intermediate ($0.3 \leq W_r \leq 1$ \AA; cyan), weak ($W_r < 0.3$ \AA; dark green), or non-detections (crosses). For \ion{Ca}{2} (\textit{bottom} row), the strong, intermediate, weak, and non-detection categories correspond to $W_r > 0.2$ \AA (blue), $0.1 \leq W_r \leq 0.2$ \AA (orange), $W_r < 0.1$ \AA (green), and non-detections (crosses), respectively. Uncertainties in the azimuthal angle, $\phi$, are indicated by dashed curves associated with each galaxy. The \textit{left} panels display all galaxies with available absorption constraints, whereas the \textit{right} panels show only those with well-determined azimuthal angles, defined by orientation uncertainties satisfying $|\Delta\phi| < 15^\circ$. The distribution of absorber strengths relative to the galaxy's major and minor axes illustrates the degree to which gas properties correlate with disk geometry (see the main text for details).}
    \label{fig:galquasdistr}
\end{figure*}

\subsection{Azimuthal distribution of metal-line absorbers}
\label{sec:azimith}

We now investigate whether the observed absorber strength exhibits any dependence on azimuthal angle.  Figure \ref{fig:galquasdistr} shows the locations of individual QSO sightlines relative to each galaxy’s semi-major and semi-minor axes, with the galaxy centered at (0,0). While the projected galaxy–QSO separation is well determined, uncertainties in the relative orientation---set by the quality of the available imaging---can be large and these are indicated by dashed curves. Crosses mark QSO sightlines with no detected absorbers, and the remaining QSO sightlines are grouped into three different categories based on the observed absorber strength.  For \ion{Mg}{2} in panels (\textit{a}) and (\textit{b}), systems are classified as strong ($W_r >$ 1 \AA), intermediate ($0.3 \leq W_r \leq 1$ \AA), weak ($W_r < 0.3$ \AA), while for \ion{Ca}{2} in panels (\textit{c}) and (\textit{d}) the strong, intermediate, and weak categories correspond to $W_r > 0.2$ \AA, $0.1 \leq W_r \leq 0.2$ \AA, $W_r < 0.1$ \AA, respectively.  The \textit{left} panels include all galaxies, while the \textit{right} panels include only those with uncertainties in the azimuthal angle $|\Delta\phi|<15^\circ$.

When considering the full sample of 87 galaxies, we find no clear relationship between the observed \ion{Mg}{2} absorption strength and the azimuthal angle (Figure \ref{fig:phidistrth}a). However, when restricting the analysis to the 71 galaxies with well-constrained azimuthal angles, there is a modest tendency for stronger \ion{Mg}{2} absorbers to lie closer to the projected major axis, with nine (five) out of 38 (33) sightlines at $<45^\circ$ from the major (minor) axis containing a strong \ion{Mg}{2} absorber of $\ewr>1$ \AA.  This is qualitatively consistent with the trend reported by \citet{Martin_2019}. To assess the significance of this apparent pattern, we perform a Kendall’s $\tau$ test and obtain $\tau = -0.1$ with a probability of no correlation $p = 0.32$. A similar analysis for the \ion{Ca}{2} measurements---based on a smaller subset of 46 galaxies with robust constraints on $\phi$---likewise reveals no statistically significant correlation between the rest-frame \ion{Ca}{2} equivalent width and the azimuthal angle.

To better understand the differences between the findings of \cite{Bordoloi_2011}, which reported excess absorption along the minor axis, the findings of \cite{Martin_2019}, which reported excess absorption along the major axis, and a lack of significant azimuthal dependence from our studies, we examine the possibility of sample selection bias.  Briefly, the galaxy sample adopted by \cite{Bordoloi_2011} focuses on disk galaxies, spanning a redshift range from $z=0.5$ to $z=0.9$ with a median of $\langle\,z\,\rangle\approx 0.75$ and a range in \mstar, from $\log\,\mstar/\msun\approx 9$ to $\log\,\mstar/\msun\approx 11.5$, while the galaxy sample adopted by \cite{Martin_2019} focuses on disk galaxies, spanning a redshift range from $z=0.136$ to $z=0.399$ with a median of $\langle\,z\,\rangle\approx 0.21$ and a range in \mstar, from $\log\,\mstar/\msun\approx 8.7$ to $\log\,\mstar/\msun\approx 10.8$ with a median of $\langle\,\log\,\mstar/\msun\,\rangle\approx 9.9$.  In comparison, galaxies extracted from the M3 sample span a redshift range from $z=0.107$ to $z=0.397$ with a median of $\langle\,z\,\rangle\approx 0.21$ and a range in \mstar, from $\log\,\mstar/\msun\approx 8.6$ to $\log\,\mstar/\msun\approx 11.2$ with a median of $\langle\,\log\,\mstar/\msun\,\rangle\approx 10.1$. At face value, the difference in the sample characteristics, coupled with the discrepancy in the azimuthal dependence of CGM \ion{Mg}{2} absorption, suggests a redshift evolution between $z\approx 0.75$ and $z\approx0.2$.  However, we caution the difficulty in interpreting stacked signals presented in \cite{Bordoloi_2011}.  A clear understanding of such discrepancies would still require targeted studies of individual galaxies at $z>0.5$.

In summary, we have carried out a detailed morphological analysis of 87 isolated galaxies drawn from the M3 halo project to assess whether the strength of \ion{Mg}{2} and \ion{Ca}{2} absorption in the inner CGM exhibits any dependence on galaxy orientation. Leveraging uniform DECaLS imaging for robust structural measurements and consistent absorption constraints from high-quality QSO spectra, we have systematically evaluated the impact of sample selection and uncertainties in inclination and position angle. We find no statistically significant correlation between absorber strength and azimuthal angle for either \ion{Mg}{2} or \ion{Ca}{2} when considering the full sample. A mild preference for stronger \ion{Mg}{2} absorption along the projected major axis emerges only when restricting to systems with well‐constrained orientations, but the trend is of low statistical significance. The \ion{Ca}{2} analysis is limited by a substantially lower covering fraction and smaller sample size, yielding no detectable azimuthal dependence. Taken together, these findings show that, within 50 kpc, the metal-enriched CGM traced by \ion{Mg}{2} and \ion{Ca}{2} does not display robust signatures of biconical outflows or co-planar inflows in our low-redshift galaxy sample. A larger sample with deeper imaging and higher-resolution spectroscopy will be essential to definitively test for subtle azimuthal trends and to reconcile the disparate results reported in previous studies across different redshifts and galaxy populations.


\section*{Acknowledgements}

RV and HWC acknowledge support from the University of Chicago Quad Faculty Research Grant, which helped initiate this project. RV is grateful for the Collegiate Research Funds that supported part of the study period. HWC acknowledges partial support from NASA ADAP grant 80NSSC23K0479.

\section*{Data Availability}

The galaxy catalog containing all relevant galaxy and absorber measurements will be made available upon request.



\bibliographystyle{mnras}
\bibliography{references} 

\clearpage



\renewcommand{\arraystretch}{1.5} 
\begin{table*}
\scriptsize
    \caption{Summary of the galaxy and absorber pair sample}
    \centering
    \begin{tabular}{lrrrrrrrrrrrr}
        \hline  
         & 
         &  \multicolumn{1}{c}{$\Delta\theta$} 
         & \multicolumn{1}{c}{$d$} 
         &  
         &\multicolumn{1}{c}{EW(H$\alpha$)}
         & \multicolumn{1}{c}{$R_h$}
         & \multicolumn{1}{c}{$r_e$} 
         & \multicolumn{1}{c}{$i$}
         & \multicolumn{1}{c}{$\phi^a$}
         & \multicolumn{1}{c}{$r_d$}
         & \multicolumn{1}{c}{$\ewr^b$}
         & \multicolumn{1}{c}{$W_r(3934)$}
         \\ 
        \multicolumn{1}{c}{ID} 
        & $ z_{\rm gal}$
        &  \multicolumn{1}{c}{($\arcsec$)} 
        & \multicolumn{1}{c}{(kpc)} 
        & \multicolumn{1}{c}{$\log\frac{\mstar}{\msun}$}  
        &\multicolumn{1}{c}{(\(\text{\AA}\))}
        & \multicolumn{1}{c}{(kpc)}
        & \multicolumn{1}{c}{(kpc)}
        & \multicolumn{1}{c}{($^\circ$)}
        & \multicolumn{1}{c}{($^\circ$)}
        & \multicolumn{1}{c}{(kpc)}
        & \multicolumn{1}{c}{(\(\text{\AA}\))} 
        & \multicolumn{1}{c}{(\(\text{\AA}\))}
        \\
        \multicolumn{1}{c}{(1)} 
        & \multicolumn{1}{c}{(2)} 
        & \multicolumn{1}{c}{(3)} 
        & \multicolumn{1}{c}{(4)}
        & \multicolumn{1}{c}{(5)} 
        & \multicolumn{1}{c}{(6)}
        & \multicolumn{1}{c}{(7)}
        & \multicolumn{1}{c}{(8)} 
        & \multicolumn{1}{c}{(9)} 
        & \multicolumn{1}{c}{(10)} 
        & \multicolumn{1}{c}{(11)}
        & \multicolumn{1}{c}{(12)} 
        & \multicolumn{1}{c}{(13)}
        \\
        \midrule
   
SDSSJ001336.14$+$141428.04$^c$ & 0.19 & 7.0 & 22.2 & 9.7 & $20.8^{+1.2}_{-1.2}$ & 143.9 & $7.8^{+0.1}_{-0.1}$ & $52.4^{+1.0}_{-1.0}$ & $34^{+1}_{-1}$ & 24.6 & $1.3^{+0.13}_{-0.13}$ & $\ldots$ \\
SDSSJ003014.16$+$011359.19 & 0.21 & 7.1 & 24.5 & 9.7 & $19.4^{+4.0}_{-4.0}$ & 142.9 & $2.1^{+0.5}_{-0.5}$ & $31.8^{+5.5}_{-6.5}$ & $88^{+2}_{-13}$ & 60.3 & $0.37^{+0.05}_{-0.05}$ & $0.16^{+0.05}_{-0.05}$ \\
SDSSJ003339.66$-$005518.39 & 0.18 & 11.0 & 32.6 & 10.1 & $<1.1$ & 172.4 & $2.8^{+0.1}_{-0.1}$ & $46.4^{+1.8}_{-2.0}$ & $80^{+2}_{-3}$ & 66.5 & $0.19^{+0.04}_{-0.04}$ & $< -0.03$ \\
SDSSJ003339.86$-$005522.39 & 0.21 & 6.3 & 21.7 & 9.8 & $46.7^{+2.9}_{-2.9}$ & 149.0 & $12.0^{+0.5}_{-0.5}$ & $71.9^{+0.8}_{-0.7}$ & $13^{+0}_{-1}$ & 25.9 & $1.05^{+0.03}_{-0.03}$ & $\ldots$ \\
SDSSJ003406.33$-$085448.74 & 0.14 & 15.6 & 38.3 & 10.6 & $<0.7$ & 228.1 & $5.6^{+0.0}_{-0.0}$ & $43.1^{+0.4}_{-0.4}$ & $72^{+0}_{-1}$ & 54.0 & $<0.14$ & $< 0.02$ \\
SDSSJ003407.78$-$085453.28 & 0.36 & 6.3 & 31.8 & 9.8 & $50.7^{+3.0}_{-3.0}$ & 134.7 & $3.9^{+0.2}_{-0.2}$ & $60.0^{+3.0}_{-3.1}$ & $50^{+3}_{-4}$ & 52.2 & $0.48^{+0.05}_{-0.05}$ & $\ldots$ \\
SDSSJ003412.85$-$010019.81 & 0.26 & 7.6 & 30.3 & 10.3 & $<3.7$ & 184.1 & $2.7^{+0.1}_{-0.1}$ & $41.4^{+2.6}_{-2.8}$ & $35^{+4}_{-5}$ & 37.0 & $0.61^{+0.06}_{-0.06}$ & $< 0.03$ \\
SDSSJ010155.80$-$084408.75 & 0.16 & 10.3 & 28.1 & 9.4 & $23.1^{+2.2}_{-2.2}$ & 134.9 & $8.8^{+0.5}_{-0.5}$ & $71.3^{+1.0}_{-0.9}$ & $35^{+0}_{-2}$ & 56.1 & $0.36^{+0.03}_{-0.03}$ & $0.14^{+0.05}_{-0.05}$ \\
SDSSJ010205.44$+$001147.92 & 0.24 & 11.6 & 43.6 & 10.4 & $15.4^{+1.0}_{-1.0}$ & 198.5 & $4.2^{+0.1}_{-0.1}$ & $21.6^{+2.2}_{-2.5}$ & $37^{+53}_{-37}$ & 47.1 & $0.57^{+0.03}_{-0.03}$ & $\ldots$ \\
SDSSJ010351.82$+$003740.75 & 0.35 & 9.7 & 47.7 & 10.5 & $19.1^{+1.4}_{-1.4}$ & 191.1 & $2.7^{+0.2}_{-0.2}$ & $28.4^{+3.7}_{-4.3}$ & $27^{+8}_{-10}$ & 49.8 & $0.38^{+0.03}_{-0.03}$ & $< 0.0$ \\
SDSSJ010508.43$-$005044.12 & 0.32 & 5.2 & 23.8 & 10.4 & $21.5^{+1.8}_{-1.8}$ & 185.1 & $5.5^{+0.1}_{-0.1}$ & $47.9^{+0.8}_{-0.7}$ & $35^{+1}_{-1}$ & 27.9 & $1.8^{+0.06}_{-0.06}$ & $< 0.01$ \\
SDSSJ012603.12$-$000834.42 & 0.13 & 12.0 & 28.0 & 9.4 & $13.4^{+1.6}_{-1.6}$ & 137.3 & $12.2^{+0.2}_{-0.2}$ & $56.6^{+0.5}_{-0.4}$ & $4^{+0}_{-1}$ & 28.2 & $0.84^{+0.08}_{-0.08}$ & $< 0.04$ \\
SDSSJ014916.68$-$081547.70 & 0.21 & 10.0 & 34.2 & 10.4 & $7.2^{+0.9}_{-0.9}$ & 200.2 & $8.5^{+0.0}_{-0.0}$ & $31.8^{+0.6}_{-0.7}$ & $55^{+1}_{-1}$ & 37.9 & $0.3^{+0.03}_{-0.03}$ & $0.29^{+0.02}_{-0.02}$ \\
SDSSJ021558.84$-$011131.26 & 0.21 & 7.9 & 27.1 & 10.2 & $<0.6$ & 177.9 & $2.1^{+0.0}_{-0.0}$ & $43.9^{+1.4}_{-1.4}$ & $28^{+3}_{-4}$ & 189.0 & $0.77^{+0.05}_{-0.05}$ & $\ldots$ \\
SDSSJ022949.97$-$074255.89 & 0.39 & 5.4 & 28.3 & 11.0 & $33.2^{+1.9}_{-1.9}$ & 276.0 & $4.5^{+0.2}_{-0.2}$ & $60.0^{+1.2}_{-1.2}$ & $37^{+1}_{-1}$ & 39.7 & $1.74^{+0.04}_{-0.04}$ & $< 0.0$ \\
SDSSJ023606.27$-$084247.01 & 0.29 & 10.4 & 45.2 & 11.0 & $11.6^{+0.8}_{-0.8}$ & 312.1 & $8.9^{+0.2}_{-0.2}$ & $70.7^{+0.8}_{-0.7}$ & $39^{+0}_{-1}$ & 45.3 & $0.62^{+0.03}_{-0.03}$ & $< 0.01$ \\
SDSSJ024425.11$-$004700.73 & 0.13 & 8.7 & 20.5 & 10.1 & $25.5^{+1.1}_{-1.1}$ & 174.4 & $8.4^{+0.1}_{-0.1}$ & $60.7^{+0.3}_{-0.4}$ & $73^{+0}_{-1}$ & 27.5 & $<0.14$ & $0.23^{+0.08}_{-0.08}$ \\
SDSSJ032232.55$+$003644.67 & 0.22 & 4.1 & 14.3 & 9.6 & $23.8^{+2.3}_{-2.3}$ & 139.3 & $6.7^{+0.7}_{-0.7}$ & $47.2^{+3.8}_{-4.2}$ & $22^{+29}_{-22}$ & 15.1 & $1.31^{+0.12}_{-0.12}$ & $0.05^{+0.01}_{-0.01}$ \\
SDSSJ075001.34$+$161301.90 & 0.15 & 8.0 & 20.4 & 8.9 & $21.3^{+1.5}_{-1.5}$ & 114.1 & $5.4^{+0.2}_{-0.2}$ & $41.4^{+3.0}_{-3.1}$ & $16^{+6}_{-7}$ & 21.5 & $0.26^{+0.08}_{-0.08}$ & $\ldots$ \\
SDSSJ075525.13$+$172825.78 & 0.25 & 12.0 & 47.2 & 10.6 & $15.4^{+0.5}_{-0.5}$ & 214.9 & $4.1^{+0.1}_{-0.1}$ & $40.5^{+1.1}_{-1.0}$ & $59^{+1}_{-2}$ & 59.2 & $0.51^{+0.02}_{-0.02}$ & $\ldots$ \\
SDSSJ075525.28$+$172833.95 & 0.31 & 3.9 & 17.6 & 9.1 & $72.2^{+3.6}_{-3.6}$ & 110.5 & $2.9^{+1.3}_{-1.3}$ & $74.9^{+8.7}_{-9.0}$ & $38^{+37}_{-38}$ & 37.2 & $0.48^{+0.02}_{-0.02}$ & $< 0.1$ \\
SDSSJ075855.37$+$181504.14 & 0.14 & 11.8  & 28.1 & 8.7 & $24.1^{+1.9}_{-1.9}$ & 105.0 & $4.7^{+0.2}_{-0.2}$ & $66.4^{+1.8}_{-1.8}$ & $22^{+1}_{-3}$ & 40.1 & $<0.1$ & $< -0.01$ \\
SDSSJ075925.72$+$104032.69 & 0.22 & 5.5 & 19.6 & 10.8 & $21.6^{+1.1}_{-1.1}$ & 262.7 & $5.9^{+0.0}_{-0.0}$ & $29.5^{+0.7}_{-0.7}$ & $86^{+0}_{-2}$ & 21.6 & $1.09^{+0.12}_{-0.12}$ & $< 0.04$ \\
SDSSJ080005.11$+$184933.28 & 0.25 & 8.2 & 32.4 & 10.0 & $31.7^{+2.4}_{-2.4}$ & 155.4 & $6.5^{+0.1}_{-0.1}$ & $70.1^{+1.4}_{-1.4}$ & $5^{+1}_{-1}$ & 33.2 & $0.3^{+0.04}_{-0.04}$ & $< 0.01$ \\
SDSSJ080113.39$+$191541.92 & 0.29 & 11.0 & 47.6 & 9.9 & $31.4^{+4.5}_{-4.5}$ & 146.1 & $2.0^{+0.7}_{-0.7}$ & $41.4^{+6.9}_{-8.0}$ & $81^{+9}_{-13}$ & 59.8 & $<0.01$ & $\ldots$ \\
SDSSJ081708.30$+$091750.33 & 0.17 & 6.3 & 18.6 & 10.6 & $15.1^{+0.2}_{-0.2}$ & 235.7 & $4.1^{+0.0}_{-0.0}$ & $39.6^{+0.7}_{-0.6}$ & $70^{+0}_{-1}$ & 23.7 & $0.7^{+0.06}_{-0.06}$ & $0.32^{+0.05}_{-0.05}$ \\
SDSSJ081731.80$+$072602.50 & 0.37 & 9.9 & 50.3 & 9.7 & $38.1^{+3.1}_{-3.1}$ & 130.6 & $2.1^{+0.2}_{-0.2}$ & $31.8^{+8.6}_{-11.7}$ & $10^{+47}_{-10}$ & 54.6 & $0.12^{+0.02}_{-0.02}$ & $< -0.0$ \\
SDSSJ082340.56$+$074751.06 & 0.19 & 12.0 & 37.4 & 10.6 & $<1.4$ & 219.9 & $6.6^{+0.2}_{-0.2}$ & $48.7^{+1.1}_{-1.1}$ & $34^{+1}_{-2}$ & 44.0 & $0.37^{+0.04}_{-0.04}$ & $\ldots$ \\
SDSSJ082806.36$+$063601.91 & 0.4 & 6.7 & 35.9 & 10.0 & $17.3^{+2.0}_{-2.0}$ & 146.3 & $4.1^{+0.2}_{-0.2}$ & $27.1^{+6.8}_{-8.9}$ & $28^{+58}_{-28}$ & 35.9 & $0.34^{+0.02}_{-0.02}$ & $< 0.03$ \\
SDSSJ084455.60$+$004718.16 & 0.16 & 11.6  & 31.0 & 10.1 & $13.8^{+1.5}_{-1.5}$ & 173.1 & $7.7^{+0.0}_{-0.0}$ & $59.3^{+0.4}_{-0.3}$ & $56^{+0}_{-1}$ & 53.9 & $0.4^{+0.05}_{-0.05}$ & $\ldots$ \\
SDSSJ085516.22$+$045232.88 & 0.33 & 9.2 & 43.3 & 9.8 & $53.1^{+3.5}_{-3.5}$ & 138.0 & $3.9^{+0.2}_{-0.2}$ & $60.0^{+3.3}_{-3.4}$ & $45^{+2}_{-4}$ & 63.5 & $0.23^{+0.02}_{-0.02}$ & $\ldots$ \\
SDSSJ090224.92$+$083045.12 & 0.12 & 10.6 & 23.0 & 9.7 & $17.0^{+1.0}_{-1.0}$ & 150.5 & $1.8^{+2.1}_{-2.1}$ & $29.5^{+14.5}_{-29.5}$ & $88^{+2}_{-59}$ & 23.8 & $0.39^{+0.09}_{-0.09}$ & $\ldots$ \\
SDSSJ090519.72$+$084914.02 & 0.15 & 2.3 & 6.1 & 8.6 & $35.1^{+1.6}_{-1.6}$ & 101.1 & $4.8^{+0.5}_{-0.5}$ & $51.7^{+4.5}_{-4.8}$ & $81^{+4}_{-4}$ & 19.7 & $0.82^{+0.01}_{-0.01}$ & $< 0.02$ \\
SDSSJ091642.92$+$014234.71 & 0.29 & 8.0 & 34.7 & 10.1 & $24.9^{+2.5}_{-2.5}$ & 157.9 & $5.3^{+0.1}_{-0.1}$ & $71.3^{+1.6}_{-1.5}$ & $18^{+1}_{-1}$ & 47.5 & $<0.04$ & $< 0.04$ \\
SDSSJ093252.24$+$073731.59 & 0.39 & 6.8 & 35.9 & 10.1 & $31.9^{+2.3}_{-2.3}$ & 152.4 & $5.7^{+0.1}_{-0.1}$ & $31.8^{+2.7}_{-3.0}$ & $90^{+0}_{-7}$ & 40.7 & $1.11^{+0.02}_{-0.02}$ & $< 0.0$ \\
SDSSJ093537.24$+$112410.65 & 0.28 & 4.5 & 19.2 & 10.2 & $17.6^{+1.3}_{-1.3}$ & 168.0 & $4.3^{+0.1}_{-0.1}$ & $38.7^{+1.6}_{-1.5}$ & $46^{+6}_{-6}$ & 22.2 & $0.79^{+0.04}_{-0.04}$ & $< 0.05$ \\
SDSSJ095051.62$+$021438.17 & 0.15 & 15.4 & 39.9 & 10.4 & $14.1^{+0.4}_{-0.4}$ & 207.2 & $7.2^{+0.1}_{-0.1}$ & $60.0^{+0.3}_{-0.3}$ & $89^{+0}_{-1}$ & 79.7 & $0.24^{+0.03}_{-0.03}$ & $\ldots$ \\
SDSSJ095605.25$+$123805.76 & 0.13 & 15.2 & 34.1 & 9.6 & $13.9^{+0.8}_{-0.8}$ & 143.5 & $4.9^{+0.1}_{-0.1}$ & $55.2^{+0.6}_{-0.5}$ & $46^{+0}_{-1}$ & 49.3 & $0.74^{+0.17}_{-0.17}$ & $0.07^{+0.02}_{-0.02}$ \\
SDSSJ100906.92$+$023557.84 & 0.25 & 8.9 & 34.9 & 10.8 & $<1.5$ & 246.9 & $4.8^{+0.1}_{-0.1}$ & $58.7^{+0.6}_{-0.6}$ & $49^{+0}_{-1}$ & 55.3 & $0.1^{+0.01}_{-0.01}$ & $< 0.02$ \\
SDSSJ102839.12$+$111828.96 & 0.24 & 7.6 & 28.5 & 9.9 & $35.1^{+2.1}_{-2.1}$ & 151.1 & $6.8^{+0.4}_{-0.4}$ & $62.6^{+2.4}_{-2.5}$ & $1^{+3}_{-1}$ & 28.5 & $<0.05$ & $\ldots$ \\
SDSSJ103251.73$+$021817.77 & 0.13 & 12.4 & 28.2 & 10.6 & $<0.4$ & 235.3 & $5.4^{+0.0}_{-0.0}$ & $45.6^{+0.1}_{-0.2}$ & $2^{+0}_{-1}$ & 28.2 & $<0.16$ & $< 0.04$ \\
SDSSJ103836.39$+$095143.66 & 0.17 & 4.9 & 14.4 & 9.1 & $18.0^{+1.1}_{-1.1}$ & 119.9 & $7.8^{+0.2}_{-0.2}$ & $65.2^{+0.8}_{-0.9}$ & $53^{+0}_{-1}$ & 29.0 & $1.04^{+0.06}_{-0.06}$ & $< -0.0$ \\
SDSSJ104533.62$+$074610.74 & 0.11 & 14.2 & 28.5 & 10.5 & $5.4^{+0.6}_{-0.6}$ & 220.0 & $22.8^{+0.1}_{-0.1}$ & $80.2^{+0.3}_{-0.3}$ & $81^{+0}_{-0}$ & 187.8 & $0.42^{+0.1}_{-0.1}$ & $\ldots$ \\
SDSSJ104839.47$+$112943.82 & 0.14 & 19.5 & 47.1 & 9.9 & $6.3^{+1.0}_{-1.0}$ & 159.5 & $4.3^{+0.2}_{-0.2}$ & $25.8^{+4.3}_{-5.0}$ & $45^{+17}_{-19}$ & 51.1 & $0.27^{+0.05}_{-0.05}$ & $\ldots$ \\
SDSSJ111508.25$+$023752.73 & 0.28 & 10.2 & 42.9 & 11.2 & $<0.8$ & 406.0 & $6.9^{+0.1}_{-0.1}$ & $32.9^{+0.9}_{-1.0}$ & $12^{+2}_{-2}$ & 45.0 & $0.44^{+0.02}_{-0.02}$ & $< -0.01$ \\
\midrule
\multicolumn{13}{l}{$^a$The orientation angle defined in Equation \ref{eq:phi} and measured using the MCMC routine described in \S\ \ref{sec:method}} \\
        \multicolumn{13}{l}{$^b$Upper limits represent 2-$\sigma$ constraints for non-detections.} \\
        \multicolumn{13}{l}{$^c$An additional bulge component is necessary to explain the galaxy morphology but only the more spatially extended disk component is listed here.} \\        \label{tab:galaxies}
    \end{tabular}
\end{table*}
\setcounter{table}{0}
\begin{table}
\scriptsize
    \caption{$-$$-$ continued}
    \centering
\begin{tabular}{lrrcrrrrrrrrr}
        \hline  
         & 
         &  \multicolumn{1}{c}{$\Delta\theta$} 
         & \multicolumn{1}{c}{$d$} 
         &  
         &\multicolumn{1}{c}{EW(H$\alpha$)}
         & \multicolumn{1}{c}{$R_h$}
         & \multicolumn{1}{c}{$r_e$} 
         & \multicolumn{1}{c}{$i$}
         & \multicolumn{1} {c}{$\phi^a$}
         & \multicolumn{1}{c}{$r_d$}
         & \multicolumn{1}{c}{$\ewr^b$}
         & \multicolumn{1}{c}{$W_r(3934)$}
         \\ 
        \multicolumn{1}{c}{ID} 
        & $ z_{\rm gal}$
        &  \multicolumn{1}{c}{($\arcsec$)} 
        & \multicolumn{1}{c}{(kpc)} 
        & \multicolumn{1}{c}{$\log\frac{\mstar}{\msun}$}  
        &\multicolumn{1}{c}{(\(\text{\AA}\))}
        & \multicolumn{1}{c}{(kpc)}
        & \multicolumn{1}{c}{(kpc)}
        & \multicolumn{1}{c}{($^\circ$)}
        &  \multicolumn{1}{c}{($^\circ$)}
        & \multicolumn{1}{c}{(kpc)}
        & \multicolumn{1}{c}{(\(\text{\AA}\))} 
        & \multicolumn{1}{c}{(\(\text{\AA}\))}
        \\
        \multicolumn{1}{c}{(1)} 
        & \multicolumn{1}{c}{(2)} 
        & \multicolumn{1}{c}{(3)} 
        & \multicolumn{1}{c}{(4)}
        & \multicolumn{1}{c}{(5)} 
        & \multicolumn{1}{c}{(6)}
        & \multicolumn{1}{c}{(7)}
        & \multicolumn{1}{c}{(8)} 
        & \multicolumn{1}{c}{(9)} 
        & \multicolumn{1}{c}{(10)} 
        & \multicolumn{1}{c}{(11)}
        & \multicolumn{1}{c}{(12)} 
        & \multicolumn{1}{c}{(13)}
        \\
        \midrule
        SDSSJ113756.75$+$085022.35 & 0.34 & 6.6 & 31.5 & 10.2 & $46.4^{+2.0}_{-2.0}$ & 163.6 & $3.1^{+0.3}_{-0.3}$ & $23.1^{+5.5}_{-7.3}$ & $8^{+41}_{-8}$ & 31.5 & $0.91^{+0.06}_{-0.06}$ & $< 0.04$ \\
SDSSJ121308.75$+$140836.90 & 0.29 & 7.1 & 30.9 & 10.4 & $24.4^{+2.0}_{-2.0}$ & 183.1 & $7.7^{+0.3}_{-0.3}$ & $75.5^{+0.8}_{-0.8}$ & $5^{+1}_{-1}$ & 32.9 & $0.23^{+0.02}_{-0.02}$ & $< 0.03$ \\
SDSSJ121347.14$+$000136.63 & 0.23 & 8.8 & 31.8 & 10.6 & $<1.4$ & 228.9 & $3.7^{+0.2}_{-0.2}$ & $11.5^{+6.5}_{-11.5}$ & $49^{+41}_{-49}$ & 36.8 & $0.54^{+0.08}_{-0.08}$ & $0.03^{+0.02}_{-0.02}$ \\
SDSSJ121640.34$+$071224.42 & 0.24 & 3.6 & 13.3 & 9.7 & $62.6^{+3.7}_{-3.7}$ & 141.8 & $10.3^{+1.2}_{-1.2}$ & $50.2^{+2.4}_{-2.5}$ & $14^{+3}_{-3}$ & 13.8 & $1.22^{+0.02}_{-0.02}$ & $\ldots$ \\
SDSSJ122933.87$+$100659.29 & 0.16 & 16.0 & 44.5 & 10.3 & $25.0^{+2.4}_{-2.4}$ & 193.0 & $13.4^{+0.3}_{-0.3}$ & $74.3^{+0.5}_{-0.4}$ & $69^{+0}_{-1}$ & 149.3 & $0.75^{+0.09}_{-0.09}$ & $< 0.02$ \\
SDSSJ132757.23$+$101135.97 & 0.26 & 6.2 & 24.3 & 9.7 & $43.7^{+3.1}_{-3.1}$ & 142.0 & $4.2^{+0.3}_{-0.3}$ & $55.2^{+2.8}_{-2.8}$ & $19^{+2}_{-4}$ & 27.4 & $0.65^{+0.04}_{-0.04}$ & $< 0.02$ \\
SDSSJ132831.54$+$075942.99 & 0.33 & 6.9 & 32.7 & 11.0 & $17.3^{+1.3}_{-1.3}$ & 304.2 & $6.1^{+0.1}_{-0.1}$ & $30.7^{+1.8}_{-2.0}$ & $66^{+6}_{-7}$ & 36.9 & $0.59^{+0.04}_{-0.04}$ & $\ldots$ \\
SDSSJ133947.23$+$093647.26 & 0.17 & 13.5 & 38.9 & 9.5 & $18.5^{+1.7}_{-1.7}$ & 135.4 & $7.2^{+0.1}_{-0.1}$ & $62.0^{+0.6}_{-0.7}$ & $34^{+0}_{-1}$ & 56.3 & $0.27^{+0.05}_{-0.05}$ & $0.18^{+0.03}_{-0.03}$ \\
SDSSJ134814.00$+$122519.78 & 0.24 & 13.2 & 49.1 & 9.6 & $28.0^{+1.5}_{-1.5}$ & 134.6 & $7.6^{+0.6}_{-0.6}$ & $71.9^{+1.4}_{-1.4}$ & $89^{+1}_{-2}$ & 163.4 & $0.26^{+0.04}_{-0.04}$ & $< -0.12$ \\
SDSSJ135344.95$+$113952.63 & 0.29 & 9.6 & 41.4 & 11.0 & $<1.6$ & 317.2 & $7.4^{+0.3}_{-0.3}$ & $48.7^{+1.1}_{-1.1}$ & $70^{+0}_{-2}$ & 60.7 & $<0.06$ & $\ldots$ \\
SDSSJ140402.15$+$102751.47 & 0.27 & 9.1 & 37.2 & 9.6 & $22.2^{+2.1}_{-2.1}$ & 134.6 & $4.6^{+0.1}_{-0.1}$ & $56.6^{+1.3}_{-1.2}$ & $51^{+1}_{-2}$ & 57.5 & $0.23^{+0.04}_{-0.04}$ & $< 0.03$ \\
SDSSJ140619.93$+$130105.24 & 0.22 & 5.0 & 17.6 & 9.8 & $26.9^{+2.2}_{-2.2}$ & 146.9 & $5.7^{+0.1}_{-0.1}$ & $55.2^{+1.1}_{-1.0}$ & $20^{+1}_{-1}$ & 19.7 & $0.96^{+0.06}_{-0.06}$ & $0.05^{+0.02}_{-0.02}$ \\
SDSSJ141453.13$+$091111.47 & 0.3 & 5.8 & 25.9 & 9.6 & $52.9^{+3.8}_{-3.8}$ & 133.3 & $4.7^{+0.4}_{-0.4}$ & $35.9^{+4.5}_{-5.0}$ & $30^{+8}_{-8}$ & 27.5 & $1.92^{+0.03}_{-0.03}$ & $< -0.01$ \\
SDSSJ141527.62$-$002640.13$^c$ & 0.14 & 18.6 & 45.5 & 10.7 & $<1.3$ & 263.9 & $7.9^{+0.0}_{-0.0}$ & $48.7^{+0.3}_{-0.3}$ & $78^{+0}_{-1}$ & 75.5 & $<0.01$ & $< -0.03$ \\
SDSSJ142114.31$+$065058.10 & 0.17 & 15.4 & 45.4 & 11.0 & $<0.8$ & 317.0 & $5.8^{+0.1}_{-0.1}$ & $28.4^{+2.5}_{-2.8}$ & $32^{+8}_{-10}$ & 47.1 & $<0.06$ & $\ldots$ \\
SDSSJ143216.97$+$095522.20 & 0.33 & 4.0 & 18.8 & 10.3 & $41.2^{+1.0}_{-1.0}$ & 170.0 & $2.9^{+0.2}_{-0.2}$ & $31.8^{+7.7}_{-9.9}$ & $73^{+15}_{-17}$ & 22.1 & $2.36^{+0.04}_{-0.04}$ & $\ldots$ \\
SDSSJ143932.21$+$000449.54 & 0.35 & 5.5 & 27.1 & 10.6 & $27.7^{+1.2}_{-1.2}$ & 205.9 & $11.6^{+0.3}_{-0.3}$ & $43.9^{+1.5}_{-1.5}$ & $88^{+2}_{-3}$ & 37.6 & $1.17^{+0.03}_{-0.03}$ & $0.13^{+0.04}_{-0.04}$ \\
SDSSJ144312.99$+$094908.40 & 0.24 & 5.2 & 20.0 & 10.0 & $20.4^{+1.7}_{-1.7}$ & 160.1 & $11.6^{+0.3}_{-0.3}$ & $83.7^{+6.3}_{-17.2}$ & $89^{+1}_{-48}$ & 22.7 & $1.08^{+0.0}_{-0.0}$ & $< 0.05$ \\
SDSSJ151228.25$-$011216.10 & 0.13 & 11.3 & 25.7 & 9.7 & $17.9^{+0.8}_{-0.8}$ & 148.5 & $6.8^{+0.1}_{-0.1}$ & $53.1^{+0.7}_{-0.7}$ & $74^{+1}_{-1}$ & 41.8 & $0.94^{+0.16}_{-0.16}$ & $\ldots$ \\
SDSSJ152424.70$+$095833.66 & 0.12 & 5.2 & 11.2 & 9.6 & $<0.7$ & 147.0 & $11.1^{+0.5}_{-0.5}$ & $50.9^{+0.7}_{-0.6}$ & $40^{+0}_{-2}$ & 14.2 & $<0.14$ & $\ldots$ \\
SDSSJ153113.01$+$091127.02 & 0.27 & 11.8 & 48.1 & 10.3 & $10.0^{+1.6}_{-1.6}$ & 181.0 & $3.6^{+0.2}_{-0.2}$ & $31.8^{+6.7}_{-8.3}$ & $61^{+21}_{-22}$ & 50.5 & $0.31^{+0.03}_{-0.03}$ & $< 0.05$ \\
SDSSJ153715.67$+$023056.40 & 0.22 & 8.7 & 30.2 & 10.0 & $45.2^{+1.5}_{-1.5}$ & 162.2 & $6.0^{+0.1}_{-0.1}$ & $24.5^{+2.0}_{-2.2}$ & $41^{+5}_{-5}$ & 31.1 & $0.8^{+0.02}_{-0.02}$ & $< 0.03$ \\
SDSSJ154956.73$+$070056.09 & 0.16 & 11.4 & 31.7 & 10.0 & $13.9^{+0.6}_{-0.6}$ & 165.2 & $7.6^{+0.1}_{-0.1}$ & $50.2^{+1.0}_{-1.0}$ & $84^{+0}_{-2}$ & 49.5 & $0.13^{+0.0}_{-0.0}$ & $0.24^{+0.03}_{-0.03}$ \\
SDSSJ155556.55$-$003615.56 & 0.3 & 10.1 & 44.9 & 9.9 & $31.1^{+2.8}_{-2.8}$ & 145.7 & $5.2^{+0.2}_{-0.2}$ & $60.0^{+1.8}_{-1.9}$ & $57^{+2}_{-2}$ & 78.5 & $<0.04$ & $< 0.07$ \\
SDSSJ204208.73$-$055224.86 & 0.11 & 8.6 & 16.8 & 9.9 & $14.7^{+1.2}_{-1.2}$ & 161.3 & $5.6^{+0.0}_{-0.0}$ & $59.3^{+0.4}_{-0.4}$ & $80^{+0}_{-1}$ & 32.6 & $0.92^{+0.2}_{-0.2}$ & $\ldots$ \\
SDSSJ204303.54$-$010139.05$^c$ & 0.24 & 13.1 & 48.6 & 10.5 & $51.4^{+0.7}_{-0.7}$ & 201.8 & $4.7^{+0.0}_{-0.0}$ & $32.9^{+0.8}_{-0.9}$ & $49^{+1}_{-2}$ & 53.0 & $1.24^{+0.05}_{-0.05}$ & $< -0.01$ \\
SDSSJ204304.35$-$010137.91 & 0.13 & 17.0 & 40.0 & 9.4 & $<0.8$ & 133.7 & $0.7^{+0.2}_{-0.2}$ & $37.8^{+15.3}_{-26.3}$ & $74^{+16}_{-24}$ & 45.4 & $<0.38$ & $< 0.01$ \\
SDSSJ204431.32$+$011304.97 & 0.19 & 7.7 & 24.7 & 10.7 & $<0.9$ & 239.1 & $6.8^{+0.7}_{-0.7}$ & $72.5^{+6.2}_{-6.3}$ & $11^{+49}_{-11}$ & 28.5 & $0.5^{+0.08}_{-0.08}$ & $< 0.0$ \\
SDSSJ210230.87$+$094121.07 & 0.36 & 4.6 & 23.0 & 10.5 & $29.6^{+2.1}_{-2.1}$ & 187.7 & $5.9^{+0.7}_{-0.7}$ & $44.8^{+3.9}_{-4.3}$ & $69^{+5}_{-6}$ & 23.7 & $0.71^{+0.04}_{-0.04}$ & $\ldots$ \\
SDSSJ212612.72$-$003819.96 & 0.21 & 10.7 & 37.0 & 10.8 & $<3.8$ & 255.8 & $6.4^{+0.0}_{-0.0}$ & $41.4^{+0.4}_{-0.4}$ & $14^{+1}_{-1}$ & 38.1 & $<0.09$ & $< 0.05$ \\
SDSSJ212938.98$-$063758.78 & 0.28 & 6.7 & 28.2 & 9.9 & $34.0^{+1.8}_{-1.8}$ & 151.1 & $3.5^{+0.1}_{-0.1}$ & $24.5^{+1.6}_{-1.7}$ & $22^{+3}_{-5}$ & 28.9 & $0.58^{+0.03}_{-0.03}$ & $0.17^{+0.03}_{-0.03}$ \\
SDSSJ221526.05$+$011353.78 & 0.32 & 10.8 & 49.9 & 10.5 & $29.5^{+1.3}_{-1.3}$ & 195.4 & $2.3^{+0.1}_{-0.1}$ & $32.9^{+2.3}_{-2.6}$ & $18^{+3}_{-4}$ & 51.0 & $0.4^{+0.05}_{-0.05}$ & $\ldots$ \\
SDSSJ223246.44$+$134655.37 & 0.32 & 8.7 & 40.3 & 11.0 & $14.7^{+0.9}_{-0.9}$ & 301.1 & $5.1^{+0.1}_{-0.1}$ & $37.8^{+1.2}_{-1.2}$ & $38^{+52}_{-38}$ & 44.7 & $1.07^{+0.06}_{-0.06}$ & $\ldots$ \\
SDSSJ223316.35$+$133315.39 & 0.21 & 9.3 & 32.2 & 10.4 & $25.2^{+1.3}_{-1.3}$ & 197.8 & $5.6^{+0.2}_{-0.2}$ & $8.1^{+9.3}_{-8.1}$ & $3^{+51}_{-3}$ & 33.5 & $1.36^{+0.06}_{-0.06}$ & $\ldots$ \\
SDSSJ223320.67$-$082539.91 & 0.36 & 10.0 & 50.0 & 10.8 & $23.8^{+2.0}_{-2.0}$ & 249.2 & $5.4^{+0.2}_{-0.2}$ & $45.6^{+2.4}_{-2.6}$ & $26^{+3}_{-5}$ & 55.6 & $1.06^{+0.03}_{-0.03}$ & $< 0.05$ \\
SDSSJ223359.74$-$003320.81 & 0.12 & 5.7 & 12.0 & 9.6 & $27.6^{+1.7}_{-1.7}$ & 146.7 & $4.6^{+0.1}_{-0.1}$ & $50.9^{+0.7}_{-0.6}$ & $44^{+0}_{-2}$ & 15.7 & $1.11^{+0.09}_{-0.09}$ & $\ldots$ \\
SDSSJ224013.49$-$083633.61 & 0.21 & 5.0 & 17.1 & 10.7 & $1.6^{+0.7}_{-0.7}$ & 239.9 & $9.7^{+0.2}_{-0.2}$ & $60.7^{+0.4}_{-0.4}$ & $89^{+0}_{-1}$ & 34.8 & $0.52^{+0.03}_{-0.03}$ & $< 0.01$ \\
SDSSJ230225.06$-$082156.62 & 0.36 & 6.9 & 34.7 & 10.7 & $62.5^{+1.4}_{-1.4}$ & 215.1 & $4.6^{+0.1}_{-0.1}$ & $58.7^{+1.6}_{-1.7}$ & $10^{+1}_{-2}$ & 35.9 & $2.02^{+0.06}_{-0.06}$ & $< -0.02$ \\
SDSSJ230845.53$-$091445.96 & 0.21 & 3.6 & 12.6 & 10.4 & $<1.7$ & 194.4 & $5.3^{+0.3}_{-0.3}$ & $41.4^{+1.4}_{-1.4}$ & $20^{+1}_{-3}$ & 13.1 & $0.43^{+0.07}_{-0.07}$ & $\ldots$ \\
SDSSJ234046.74$+$145359.14 & 0.38 & 8.6 & 44.5 & 10.9 & $<1.6$ & 268.9 & $5.1^{+0.1}_{-0.1}$ & $53.8^{+1.0}_{-0.9}$ & $17^{+0}_{-2}$ & 47.9 & $<0.04$ & $\ldots$ \\
SDSSJ234949.41$+$003542.32 & 0.28 & 7.6 & 31.9 & 10.5 & $13.0^{+1.2}_{-1.2}$ & 205.4 & $3.8^{+0.1}_{-0.1}$ & $60.7^{+1.7}_{-1.8}$ & $5^{+1}_{-2}$ & 32.0 & $0.35^{+0.06}_{-0.06}$ & $0.26^{+0.03}_{-0.03}$ \\
SDSSJ235735.97$+$003112.94 & 0.15 & 17.3 & 43.9 & 10.8 & $8.1^{+0.3}_{-0.3}$ & 286.4 & $11.7^{+0.0}_{-0.0}$ & $70.1^{+0.0}_{-0.0}$ & $70^{+0}_{-1}$ & 125.8 & $0.48^{+0.06}_{-0.06}$ & $< -0.03$ \\
        \midrule
        \multicolumn{13}{l}{$^a$The orientation angle defined in Equation \ref{eq:phi} and measured using the MCMC routine described in \S\ \ref{sec:method}} \\
        \multicolumn{13}{l}{$^b$Upper limits represent 2-$\sigma$ constraints for non-detections.} \\
        \multicolumn{13}{l}{$^c$An additional bulge component is necessary to explain the galaxy morphology but only the more spatially extended disk component is listed here.} \\
    \end{tabular}
\end{table}
\label{lastpage}
\end{document}